\documentclass[showpacs,preprintnumbers,amsmath,amssymb, nofootinbib]{revtex4}
\usepackage{fontenc}
\usepackage{graphicx}

\usepackage[dvips]{hyperref}

\usepackage{amsmath}

\begin{document}

\title{Hawking Radiation for Scalar and Dirac Fields in Five Dimensional Dilatonic Black Hole via Anomalies}
\author{Ram\'{o}n Becar}
\email{rbecar@uct.cl}
\affiliation{Departamento de Ciencias Matem\'{a}ticas y F\'{\i}sicas, Universidad Cat\'{o}lica de Temuco, Montt 56,
Casilla 15-D, Temuco, Chile.}
\author{P. A. Gonz\'{a}lez}
\email{pgonzalezm@ucentral.cl}
\affiliation{Escuela de Ingenier\'{i}a Civil en Obras Civiles. Facultad de Ciencias F\'{i}sicas y Matem\'{a}ticas, \\Universidad Central de Chile, Avenida Santa Isabel 1186, Santiago, Chile}
\affiliation{Universidad Diego Portales, Casilla 298-V, Santiago, Chile.
}
\date{\today}

\begin{abstract}

We study massive scalar fields and Dirac fields propagating in a five dimensional dilatonic black hole background. We expose that for both fields the physics can be describe by a two dimensional theory, near the horizon. Then, in this limit, by applying the covariant anomalies method we find the Hawking flux by restoring the gauge invariance and the general coordinate covariance, which coincides with the flux obtained from integrating the Planck distribution for fermions.

\end{abstract}
\maketitle

\section{Introduction}

The Hawking radiation is an important quantum effect in black hole physics. At quantum level, the black holes emit radiation with a temperature
given by $h/8 \pi k_{B}GM$ \cite{Hawking:1974sw}, while classically nothing can escape from them. This radiation is essentially thermal and
they slowly evaporate through emitting quanta. The Hawking radiation has been
considered as a main tool to understand the quantum nature of
gravity and it is determined by universal properties of the event horizon. The originated thermal radiation at
the event horizon is emitted to surrounding spacetime
which in the semiclassical approach for a black
hole means that it slowly looses its mass and eventually
evaporates. At the event horizon, the Hawking radiation is in fact
a blackbody radiation. However, this radiation still has to traverse
a non-trivial curved spacetime geometry before it reaches a distant
 observer who  detects it. The surrounding spacetime
therefore works as a potential barrier for the radiation giving a
deviation from the blackbody radiation spectrum, seen
by an asymptotic observer \cite{Harmark:2007jy,Maldacena:1996ix}. Apart from Hawking's original derivation \cite{Hawking:1974sw, Hawking:1974rv}, which calculates the
Bogoliubov coefficients between in and out states for a body collapsing to form a black hole,
there are other approaches to obtain the Hawking radiation, one
of them, was the development by Christensen and Fulling
\cite{Christensen:1977jc}, which showed that Hawking radiation in
(1+1)-dimensional Schwarzschild background metric can be derived
from the trace anomaly or conformal anomaly, which arises from the
renormalization of the quantized theory. Another derivation, known as tunneling method, was proposed in Ref.~\cite{Parikh:1999mf}, here the imaginary part of the action for the (classically forbidden) process of s-wave emission across the horizon is related to the Boltzmann factor for emission using the null geodesic equation and Hamilton's equation for outgoing particles. 

A recent approach was developed by Robinson and Wilczek, Ref.~\cite{Robinson:2005pd}. Here, the authors
showed, that in order to avoid a breakdown of general covariance at the quantum level, the total flux in each outgoing partial wave of a quantum field
in a black hole background must be equal to that of (1+1) dimensional black body at the Hawking temperature,
considering consistent gravitational anomalies in Schwarzschild type black holes, which was extended for charged black holes considering consistent gravitational and gauge anomalies, Ref.~\cite{Iso:2006wa}, and for spherically symmetric black holes, Ref.~\cite{Vagenas:2006qb}.
The alternative covariant formulation of the method was developed in Refs.~\cite{Banerjee:2007qs, Banerjee:2007uc, Banerjee:2008az, Banerjee:2008wq, Gangopadhyay:2007hr, Gangopadhyay:2008zw}. The method has acquired a growing interest and has been applied for several black holes. See Refs.~\cite{Setare:2006hq, Wu:2007sw, Miyamoto:2007ue, Chen:2007pp, Das:2007ru, Kui:2007dy, Iso:2006xj, Iso:2006ut, Wei:2009kg, Wei:2009zt} for asymptotically flat black holes, Refs.~\cite{Papantonopoulos:2008wp, Porfyriadis:2009zs} for asymptotically AdS black holes and Refs.~\cite{Kim:2007ge, Becar:2010z1} for acoustic geometries. Also, see, Refs.~\cite{ Li:2009zzw, Iso:2007nf, Morita:2009mt, Wu:2008yx, Bonora:2008he, Bonora:2008nk, Bonora:2009tw, Li:2010eca} for other metrics, generalizations and modifications of the method and Ref.~\cite{Jiang:2010ma} to consider back reaction. These last two approaches, had been connected through chirality, Ref.~\cite{Banerjee:2008sn}. See, also Ref.~\cite{Akhmedova:2008au}, where the authors comment on anomaly versus WKB/tunneling methods to calculate Unruh radiation.
 
The purpose of this work is to consider an interesting class of black
hole known as dilatonic black hole \cite{Shin:2007gz, Peng:2007pk, Jiang:2007mi}, which arise from low energy effective field theory
and have qualitatively different properties from those in ordinary Einstein
gravity. The stability of five dimensional dilatonic black holes against scalar and fermionic perturbation has been studied in Refs.~\cite{Becar:2007hu, Becar:2010z2}. While, in the first article the stability depend on the non-minimal parameter present in the quasinormal frequencies in the second one is always stable. 

Here, we study the thermal properties of these black holes via the covariant anomalous point of view for gauge and gravitational anomalies using only covariant boundary conditions at the event horizon, Ref.~\cite{Gangopadhyay:2008zw} considering both scalar fields and Dirac fields, Ref.~\cite{Li:2010eca}. In order to avoid a general covariance and gauge invariance breakdown at the quantum level, the total flux in each outgoing partial wave of a quantum field must be equal to that of a (1+1) dimensional blackbody at the Hawking temperature with the appropriate chemical potential.

The outline of this paper is as follows: in Sec. II, we specify
the geometry of the dilatonic black hole in five-dimensions. In Sec. III, we consider massive scalar
fields non minimally coupled to the curvature and we study the action near the event horizon. In Sec. IV, we consider massive fermionic fields and we study the action near the event horizon. By doing so, we demonstrate that for both fields the physics can be describe by a two dimensional theory, near the horizon.
Then, in Sec. V, we find the current flux and the energy-momentum flux. Finally, in Sec. VI, we conclude by comparing the fluxes obtained in the previous section with the flux of Hawking radiation for fermions.


\section{Five Dimensional Dilatonic Black Hole}

There has been a growing interest in five-dimensional dilatonic black holes through the years, since it is believed that these black holes could shed some light
on the fundamental problem of the microscopic origin on the
Bekenstein-Hawking entropy. The area-entropy relation $S_{BH}=A/4$ was
obtained for a class of five-dimensional extremal black holes in Type II
string theory, using D-brane techniques \cite{Strominger:1996sh}. Also, in
\cite{Teo:1998kp}, the U-duality that exists between the five dimensional
black holes an the two dimensional charged black holes \cite{McGuigan:1991qp}
was used to microscopically compute the latter entropy. The metric for
dilatonic black holes in five dimensions can be written as \cite{McGuigan:1991qp,Teo:1998kp}
\begin{equation}\label{eq1}
ds^{2}=-\left(1-\frac{r_0^2}{r^2}\right)\left(1+\frac{r_0^2\sinh^2\alpha}{r^2}\right)^{-2}dt^2+\left(\frac{r^{2}}{r_0^2}-1\right) ^{-1}dr^{2}+r_{0}^{2}d\Omega_3 ^2~.
\end{equation}
This metric is the product of two completely decoupled parts, namely, an
asymptotically flat two-dimensional which describes a two-dimensional
charged dilatonic black hole and a three sphere with constant radius, $r_0$. This
statement can be directly show if we apply in $(r,t)$ sector the
transformation defined by
\begin{equation}\label{eq2}
e^{Qx}=2\left(\frac{r^2}{r_0^2}+sinh^2\alpha\right)\left(m^2-q^2\right)^\frac{1}{2}~,
\end{equation}
where $m$ and $q$ are related to the mass and charge of the dilatonic black hole. So, the Eq.~(\ref{eq1}) read as follow
\begin{equation}\label{eq3}
ds^{2}=-N^{2}dt^2+N^{-2}dx^2+r_0^2d\Omega^{2}_{3}~,
\end{equation}
\begin{equation}\label{eq7}
\nonumber e^{-2\left(\phi-\phi_{0}\right)}=e^{Qx}~,\quad A=\sqrt{2}Qqe^{-Qx}dt~,
\end{equation}
with
\begin{equation}\label{eq4}
N^{2}=1-2me^{-Qx}+q^2e^{-2Qx}~,
\end{equation}%
where $Q$ is a positive
constant determined by the central charge and $A$ is the gauge potential. Performing the change of variables $y=e^{Qx}$, the metric (\ref{eq3}) can be written as
\begin{equation}\label{metric}
ds^{2}=-\frac{(y-y_{1})(y-y_{2})}{y^{2}}dt^2+\frac{
dy^{2}}{Q^{2}\left( y-y_{1}\right) \left( y-y_{2}\right) }+r_{0}^{2}d\Omega^{2}_{3}\
\end{equation}
\begin{equation}
\nonumber e^{-2\left(\phi-\phi_{0}\right)}=y~,\quad A=\frac{\sqrt{2}Qq}{y}dt~.
\end{equation}
The asymptotic region is given by $y=0$, while the two horizons are given by $y_{1,2}=\left(m\mp\sqrt{m^2-q^2}\right)$. Therefore, the above metric can be expressed as
\begin{equation}\label{metricequation}
ds^2=-f\left(y\right)dt^2+h\left(y\right)^{-1}dy^2+r_{0}^{2}d\Omega^{2}_{3}~,
\end{equation}
where
\begin{equation}
f\left(y\right)= \frac{(y-y_{1})(y-y_{2})}{y^{2}}~,\quad h\left(y\right)= Q^{2}\left( y-y_{1}\right) \left( y-y_{2}\right)~.
\end{equation}


\section{Scalar fields. Dimensional reduction}

We consider complex scalar field coupled non-minimally to the curvature in the background of a
five dimensional dilatonic black hole, (\ref{metric}). The action for the scalar field $\psi$ can be written as
\begin{eqnarray}\label{eq8}
\nonumber S\left[\psi\right]&=&\frac{1}{2}\int d^5y\sqrt{-g}\psi^*\left(\square
+m^2+\xi R\right)\psi~,\\
&=&\frac{1}{2}\int dtdy\int d\phi\int d\theta\frac{d\chi r_0^3sin^2\chi sin\theta}{Qy}\psi^*\left[\frac{-1}{f}\partial
_t^2+yQ\partial_{y}\left(\frac{h\left(y\right)}{yQ}\partial
_{y}\right)+m^2+\xi R+\frac{1}{r_0^2}\nabla_{(S^3)}^2\right]\psi~,
\end{eqnarray}
where $m$ is the mass of the scalar field, $\xi$ is a parameter from the non-minimal coupling, $\nabla^{2}_{(S^3)}$ is the Laplace-Beltrami operator on the $S^3$ sphere and R is
the scalar curvature given by
\begin{equation}\label{R}
R=\frac{6}{r_0^2}-Q^2y f'\left(y\right)-Q^2y^2f''\left(y\right)~,
\end{equation}
with prime denoting differentiation with respect to $y$. Performing the following decomposition of $\psi$ in terms of spherical harmonics
\begin{equation}
\psi=\sum_{nlm}\psi_{nlm}\left(t,y\right)
Y_{nlm}\left(\chi, \theta,\phi\right)~,
\end{equation}
where $(n, l, m)$ is the collection of angular quantum numbers, $Y$ is a normalizable harmonic
function on $S^{3}$, i.e., it
satisfies the equation $\nabla^{2}_{(S^3)}Y=\alpha Y$, which in terms of the
coordinates in $S^{3}$ can be written as
\begin{equation}
csc^{2}\chi \left( \frac{\partial }{\partial \chi }\left( \sin
^{2}\chi
\frac{\partial Y}{\partial \chi }\right) +csc^{2}\theta \left( \frac{%
\partial }{\partial \theta }\left( \sin ^{2}\theta \frac{\partial Y}{%
\partial \theta }\right) \right) +csc\theta \frac{\partial ^{2}Y}{\partial
\phi ^{2}}\right) =\alpha Y^{(nlm)}~,  \label{eigeneq}
\end{equation}
with
\begin{equation}
\alpha =-n(n+2),\text{ }\left| m\right| \leq l\leq n=0,1,2,...~,
\label{eigenvalues}
\end{equation}
and
\begin{equation}
Y^{(nlm)}\left( \chi ,\theta ,\phi \right) =\left( \frac{%
2^{2l+1}(n+1)(n-l)!l!^{2}}{\pi (n+l+1)!}\right) \sin ^{l}\chi
C_{n-l}^{(l+1)}\left( \cos \chi \right) Y^{(lm)}(\theta ,\phi )~,
\label{egnefunciton}
\end{equation}
here $C_{n-l}^{(l+1)}\left( \cos \chi \right) $ are the
Gegenbauer polynomials \cite{Abra,vernon},
$Y^{(lm)}(\theta ,\phi )$ are the $S^{2}$ scalar harmonics and
the coefficient is chosen to normalize the harmonics. Integrating on $\theta$, $\phi$ and  $\chi$ and by using the normalization and orthogonality of the spherical harmonic, we find that
the action can be written as
\begin{equation}\label{eq8}
\nonumber S\left[\psi\right]=\frac{1}{2}\sum_{nlm}\int dtdy \frac{r_0^3}{Qy}\psi^*_{nlm}(t,y)\left[\frac{-1}{f}\partial
_t^2+yQ\partial_{y}\left(\frac{h\left(y\right)}{yQ}\partial
_{y}\right)+m^2+\xi R+\frac{\alpha}{r_0^2}\right]\psi_{nlm}(t,y)~.
\end{equation}

Besides, with order to study the physic near to the event horizon results helpful to work in tortoise coordinate $y_{\ast}$, which is defined by
\begin{equation}\label{tortoise}
\frac{\partial y_{\ast}}{\partial y}=\frac{1}{\sqrt{f(y)h(y)}}~,
\end{equation}
in this way the action in tortoise coordinate is
\begin{equation}\label{actiontortoise}
S_*\left[\psi\right]=\frac{1}{2}\sum_{nlm}\int dtdy_{\ast}f\left(y\left(y_{\ast}\right)\right)r_0^3\psi^*_{nlm}\left[\frac{-1}{f}\partial
_t^2+yQ\partial_{y}\left(\frac{h\left(y\right)}{yQ}\partial
_{y}\right)+m^2+\xi R+\frac{\alpha}{r_0^2}\right]\psi_{nlm}~.
\end{equation}
Considering the Taylor's expansion of $f\left(y\right)$ and $h\left(y\right)$
\begin{equation}
f\left(y\right)=f\left(y_H\right)+\sum_{n=1}^\infty\frac{f^{(n)}\left(y_H\right)}{n!}(y-y_H)^n~,
\end{equation}
\begin{equation}
h\left(y\right)=h\left(y_H\right)+\sum_{n=1}^\infty\frac{h^{(n)}\left(y_H\right)}{n!}(y-y_H)^n~,
\end{equation}
which at the horizon can be described by the first term of the summatory
\begin{equation}\label{aproxf}
f\left(y\right)\approx f^{'}\left(y_H\right)(y-y_H)=k_1(y-y_H)~,
\end{equation}
\begin{equation}\label{aproxh}
h\left(y\right)\approx h^{'}\left(y_H\right)(y-y_H)=k_2(y-y_H)~,
\end{equation}
and replacing Eq.~(\ref{aproxf}) and Eq.~(\ref{aproxh}) in Eq.~(\ref{tortoise}), we obtain
\begin{equation}
y\left(y_{\ast}\right)=Ae^{ky_{\ast}}+y_{H}~,
\end{equation}
where $k$ is the surface gravity defined by $k=\frac{1}{2}\sqrt{f'(y_{H})h'(y_{H})}$. So, at the event horizon we have
\begin{equation}
f\left(y\left(y_{\ast}\right)\right)\approx Ak_1e^{2ky_{\ast}}~,\quad \qquad
h\left(y\left(y_{\ast}\right)\right)\approx Ak_2e^{2ky_{\ast}}~.
\end{equation}
Note that at event horizon $(y_{\ast}\rightarrow-\infty)$, $f\left(y\left(y_{\ast}\right)\right)$ and $h\left(y\left(y_{\ast}\right)\right)$ act
as suppression factors vanishing exponentially fast, in this limit. By analyzing the terms of the action (\ref{actiontortoise}), we can see that near the event horizon the term coming from the Laplacian on unit three sphere and the mass term are suppressed by the factor $f\left( y\left( y_{\ast}\right)\right)$.

On the other hand, the part of the action~(\ref{actiontortoise}), which describes the non-minimal coupling of the scalar field to the curvature can be written as
\begin{equation}\label{Sast}
S_{c_{\ast}}\left[\psi\right]=\frac{1}{2}\sum_{nlm}\int dtdy_{\ast}f\left(y\left(y_{\ast}\right)\right)\xi\psi^*_{nlm}(t,y\left(y_{\ast}\right))\left[\frac{6}{r_{0}^{2}}-Q^{2}y\left(y_{\ast}\right)f'\left(y\left(y_{\ast}\right)\right)-Q^{2}y^{2}\left(y_{\ast}\right)f''\left(y\left(y_{\ast}\right)\right)\right]\psi_{nlm}(t,y\left(y_{\ast}\right))~,
\end{equation}
where we have used $h\left(y\right)=Q^{2}y^{2}f\left(y\right)$. 
However,  as the event horizon is not a curvature singularity, the non-minimal coupling does not affect the dimensional reduction procedure since the part of the action which describes the non-minimal coupling vanishes in the region near the event horizon. So, by the latter arguments the total action near
the event horizon is given by
\begin{equation}
S\left[\psi\right]=\frac{1}{2}\sum_{nlm}\int dtdyr_0^3\sqrt{-g}\psi^*_{nlm}(t,y)\left[\frac{-1}{f}\partial_t^2+h\partial_y^2+\frac{\left(fh\right)_{,y}}{2f}\partial_{y} \right]\psi_{nlm}(t,y)~.
\end{equation}
Thus, the physics near the horizon can be described by an infinite collection of $1+1$ dimensional complex scalar fields in the following background
\begin{equation}
ds^2=-f\left(y\right)dt^2+h\left(y\right)^{-1}dy^2~,
\end{equation}
\begin{equation}
A_{t}=\frac{\sqrt{2}Qq}{y}~,
\end{equation}
which correspond at Mcguigan, Nappi and Yost metric \cite{McGuigan:1991qp} and $\Phi=r_0^3$
is the two-dimensional dilaton field, which does not contribute to the flux.


\section{Dirac Fields. Dimensional reduction}

The formalism of differential forms  lead us to use the
fuenfbein 1-form $e^{a}=e_{\mu}^{a}dx^{\mu}$ and the spin connection 1-form
$\omega^{ab}=\omega _{\mu}^{ab}dx^{\mu}$\footnote{%
Latin indices label the components with respect to a local Lorentz frame and
Greek indices refers to the coordinate frame.}, we take our local coordinates to be $x^{\mu }=t,y,\theta, \phi, \chi$.
However, by simplicity in the notation, it is standard to work with a dual spin
connection
\begin{equation}
\omega ^{a}=-\frac{1}{2}\epsilon ^{abc}\omega _{bc}~.
\end{equation}%
So, the torsion $T^{a}$ and the curvature $R^{a}$ are given by
\begin{equation}
T^{a}=de^{a}+\epsilon_{bc}^{a}\omega^{b}e^{c}~,
\end{equation}%
\begin{equation}
R^{a}=d\omega^{a}+\frac{1}{2}\epsilon_{bc}^{a}\omega^{b}\omega ^{c}~.
\label{Curvature}
\end{equation}

For the metric, Eq.~(\ref{metricequation}), we choose the following fuenfbein
\begin{equation}
e^{0}=-\sqrt{f(y)}dt~,\quad e^{1}=\frac{1}{\sqrt{h(y)}}dr~,\quad e^{2}=r_{0}d\chi~,\quad e^{3}=r_0sin\chi d\theta~,\quad e^{4}=r_0sin\chi sin\theta d\phi~,
\end{equation}
and for vanishing torsion the connection is given by 
\begin{equation}
\omega^1_0=-\frac{1}{2f(y)}f'(y)dy~,\quad\omega^3_2=-cos\chi d\theta~,\quad \omega^4_2= -cos\chi sin\theta d\phi~,\quad \omega^4_3= -cos\theta d\phi~.  
\end{equation}

The action for Dirac fields, propagating in a five dimensional dilatonic black hole can be written as 
\begin{eqnarray}\label{actiondiracfield}
S&=&\int{d^5x\sqrt{-g}\Psi^{\dag}\gamma^{0}\left[\gamma^{a}e^{\mu}_{a}\left(\partial_{\mu}+\frac{1}{8}\left[\gamma^{b},\gamma^{c}\right]\omega_{\mu bc}\right)+m\right]\Psi}\\
&=&\int{d^5x\sqrt{-g}\Psi^{\dag}\gamma^{0}\left[\gamma^{0}e^{t}_{0}\left(\partial_t -\frac{1}{4}\omega_{10t}\gamma^1\gamma^0\right)+\gamma^1e^y_1\partial_r+\frac{1}{r_0}e^i_m\gamma^m\left(\partial_i-\frac{1}{4}\omega_{bci}\gamma^b\gamma^c\right)+m\right]\Psi}~,
\end{eqnarray}
where $e^{\mu}_a$ is the inverse fuenfbein. Now, using the descomposition
\begin{equation}
\psi=\frac{1}{f^{\frac{1}{4}}}\sum_{nlm}\phi_{nlm}\left(t,y\right)\otimes \eta_{nlm}\left(\theta,\phi,\chi \right)~,
\end{equation}
where $\phi_{nlm}\left(t,y\right)$ is a spinor and $\eta_{nlm}\left(\theta,\phi,\chi \right)$ are the eigenfunctions of the Dirac operator on three sphere. The Dirac operator, \cite{Camporesi:1995fb}, is given by 
\begin{equation}
e^i_m\gamma^m\left(\partial_i-\frac{1}{4}\omega_{bci}\gamma^b\gamma^c\right)\eta_{nlm}\left(\theta,\phi,\chi \right)=\pm i\left(n+\frac{3}{2}\right)\eta_{nlm}\left(\theta,\phi,\chi \right)~,
\end{equation}
where $\pm i\left(n+\frac{3}{2}\right)$, are the eigenvalues of the Dirac operator on the three sphere.
Thus, using the tortoise coordinate transformation, given by Eq.~(\ref{tortoise}) allows us to write the action near the horizon as 
\begin{equation}
S\left[\psi\right]=\sum_{nlm}\int dtdy_*\phi^{\dag}_{nlm}(t,y)\gamma^0\left[\gamma^0\partial_t-\frac{1}{4f}(\partial_{y_*} f)\gamma^1+\gamma^1\partial_{y_*} \right]\phi_{nlm}(t,y)~,
\end{equation}
which in $(y,t)$ sector can be written as following form
\begin{equation}
S\left[\psi\right]=\sum_{nlm}\int dtdy\phi^{\dag}_{nlm}(t,y)\gamma^0\left[\frac{\gamma^0}{\sqrt{fh}}\partial_t-\frac{1}{4f}(\partial_{y}f)\gamma^1+\gamma^1f\partial_{y} \right]\phi_{nlm}(t,y)~.
\end{equation}

Thus, the physics near the horizon can be described by an infinite collection of $1+1$ dimensional  fermionic fields in the following background
\begin{equation}
ds^2=-f\left(y\right)dt^2+h\left(y\right)^{-1}dy^2~,
\end{equation}
\begin{equation}
\nonumber A_{t}=\frac{\sqrt{2}Qq}{y}~,
\end{equation}
which correspond at Mcguigan, Nappi and Yost metric \cite{McGuigan:1991qp}.

\section{Anomalies and Hawking fluxes}
In the previous sections we have reduced the five dimensional theory at  (1+1)-dimensional background. However,  at the horizon, outgoing modes of the (1+1)- dimensional fields behave as right moving modes, while ingoing modes as left moving modes. So, if we neglect the ingoing modes in the region near the horizon, because they can not classically affect physics outside the horizon, then the effective two-dimensional theory becomes chiral and two-dimensional chiral theory contains gauge and gravitational anomalies.

In this section we will obtain the fluxes of the current and energy-momentum tensor, considering the covariant anomalous
point of view for gauge and gravitational anomalies proposed in Ref.~\cite{Gangopadhyay:2008zw}. Where, the author considered
only covariant boundary conditions at the event horizon.

\subsection{Gauge anomaly}
In order to apply the method mentioned above and study the region near the horizon, the spacetime is divided into two regions, due to the effective theory is different. Being
one of them the region outside the event horizon and the other near the horizon. In this way, the current
$J^{\mu}$, which outside the horizon is denoted by $J_{\left(o\right)}^{\mu}$ is anomaly free and satisfies the
conservation law
\begin{equation}\label{eq10}
\nabla_{\mu}J_{\left(o\right)}^{\mu}=0~,
\end{equation}
which yields
\begin{equation}\label{eq14}
J_{\left(o\right)}^y=\frac{c_o}{\sqrt{-g}}~,
\end{equation}
where $c_0$ is an integration constant. However, near the horizon
satisfies
\begin{equation}\label{eq11}
\nabla _{\mu }J_{\left(H\right)}^{\mu}=-\frac{e^{2}}{4\pi}\overline{\epsilon}^{\rho\sigma}F_{\rho\sigma}=\frac{e^{2}}{2\pi\sqrt{-g}}
\partial_{y}A_{t}~,
\end{equation}
where, the right side of the above equation corresponds to the covariant form of the two dimensional abelian anomaly and $\overline{\epsilon}^{\mu\nu}=\frac{\epsilon^{\mu\nu}}{\sqrt{-g}}$ and $\overline{\epsilon}_{\mu\nu}=\sqrt{-g}\epsilon _{\mu\nu}$ are
two dimensional antisymmetric tensors for the upper and lower cases with $\epsilon^{ty}=\epsilon _{ty}=1$. The solution of this differential equation is given by
\begin{equation}\label{eq15}
J_{\left(H\right)}^y=\frac{1}{\sqrt{-g}}\left(c_H+\frac{e^2}{2\pi}\left[A_t\left(y\right)-A_t\left(y_H\right)\right]\right)~,
\end{equation}
where $c_H$ is an integration constant. On the other hand, $J^y\left(y\right)$ and $\nabla_{\mu}J^{\mu}$ can be written as
\begin{equation}
J^y\left(y\right)=J_{\left(o\right)}^y\left(y\right)\Theta\left(
y-y_{H}-\epsilon\right)+J_{\left(H\right)}^y\left(y\right)H\left(y\right)~,
\label{eq16}
\end{equation}
where, the outside region near the horizon is defined for: $y_H \le y \le  y_H + \epsilon$ and the outside region for: $y_H +\epsilon \le y \le  \infty$.
\begin{equation}\label{eq17}
\nabla_{\mu}J^{\mu}=
\left[\sqrt{-g}\left(J_{\left(o\right)
}^y\left(y\right)-J_{\left(H\right)}^y\left(y\right)\right)+\frac{e^2}{2\pi}
A_t\left(y\right)\right]\delta\left(y-y_{H}-\epsilon\right)
+\partial _{y}\left(\frac{e^{2}}{2\pi}A_{t}\left(y\right)H\left(
y\right)\right)~,
\end{equation}
where $H\left( y\right) =1-$ $\Theta \left( y-y_{H}-\epsilon \right)$. If we require an anomaly free and gauge invariant theory, then the term in the total derivative is canceled by quantum effects of classically irrelevant ingoing modes and the vanishing of the Ward
identity under gauge transformation implies that
\begin{equation}
J_{\left( o\right) }^{y}\left( y\right) -J_{\left( H\right)
}^{y}\left(y\right) +\frac{e^{2}}{2\pi \sqrt{-g}}A_{t}\left( y\right) =0~.
\label{eq18}
\end{equation}
Then, replacing Eq.~(\ref{eq14}) and Eq.~(\ref{eq15}) in the above equation yields
\begin{equation}
c_{o}=c_{H}-\frac{e^{2}}{2\pi }A_{t}\left( y_{H}\right)~,  \label{eq19}
\end{equation}%
where the coefficient $c_{H}$ vanishes by requiring that the covariant current $J_{\left( H\right) }^{y}\left(y\right)$ vanishes at the horizon. Therefore, the charge flux
corresponding to $J^{y}\left( y\right) $ is given by
\begin{equation}\label{c0}
c_0=\sqrt{-g}J_{\left(o\right)}^{y}\left(y\right)=-\frac{e^2}{2\pi}%
A_{t}\left(y_{H}\right)=-\frac{e^{2}}{2\pi y_{H}}\sqrt{2}Qq~.
\end{equation}
\subsection{Gravitational anomaly}
Such as mentioned the effective quantum theory is chiral near the horizon and chiral theories contain gravitational anomalies. In consequence, the divergence of  energy-momentum tensor in not vanished and this is quantum mechanically violated due to the chiral anomaly. Therefore, a real energy-momentum flux it is needed as a compensating object. The main idea of the method, consist in the divergence of this flux will be canceled the anomaly at the horizon. 

So, the energy-momentum tensor in the outside region satisfies the conservation law
\begin{equation}
\nabla _{\mu}T_{\left(o\right)\nu}^{\mu}=F_{\mu \nu }J_{\left(o\right)
}^{\mu }~,\label{eq21}
\end{equation}
outside the horizon. Which leads to
\begin{equation}
T_{\left(o\right)t}^{y}=\frac{1}{\sqrt{-g}}\left(a_{o}+c_{o}A
_{t}\left(y\right)\right)~,\label{eq28}
\end{equation}
where $a_{0}$ is an integration constant and having used $F_{yt}=\partial _{y}A_{t}$ and Eq.~(\ref{eq14}). However, once near the horizon the divergence of energy-momentum tensor satisfies
\begin{equation}
\nabla _{\mu }T_{\left( H\right) \nu }^{\mu }=F_{\mu \nu }J_{\left( H\right)
}^{\mu }+\frac{1}{96\pi }\overline{\epsilon }_{\nu \mu }\partial ^{\mu
}R=F_{\mu \nu }J_{\left( H\right) }^{\mu }+\mathcal{A}_{\nu }~,\label{eq22}
\end{equation}
where $R$ is the Ricci scalar, given by
\begin{equation}
R=-\frac{hf^{\prime \prime }}{f}-\frac{f^{\prime }h^{\prime }}{2f}+\frac{1}{2%
}\frac{\left( f^{\prime }\right) ^{2}h}{f^{2}}~,  \label{eq23}
\end{equation}
for the metric (\ref{metricequation}) and the anomaly is purely timelike with
\begin{equation}
\mathcal{A}_{y}=0~,\qquad
\mathcal{A}_{t}=\frac{1}{\sqrt{-g}}\partial_{y}N_{t}^{y}~,\label{eq25}
\end{equation}
where
\begin{equation}
N_{t}^{y}=\frac{1}{96\pi }\left( -hf^{\prime \prime }-\frac{f^{\prime
}h^{\prime }}{2}+\frac{\left( f^{\prime }\right) ^{2}h}{f}\right)~.
\label{eq26}
\end{equation}
In this way, the Eq.~(\ref{eq22}) can be written as
\begin{equation}
\partial _{y}\left(\sqrt{-g}T_{\left( H\right)t}^{y}\right)=
\partial _{y}\left(\frac{e^{2}}{2\pi}\left[\frac{1}{2}A
_{t}^{2}\left(y\right)-A_{t}\left(y_{H}\right)A
_{t}\left(y\right)\right]+N_{t}^{y}\left(y\right)\right)~,\label{eq29}
\end{equation}
where the first term comes from  the gauge anomaly and the second of the gravitational anomaly. Here, we have used Eq.~(\ref{eq15}) with $c_{H}=0$, in order for the covariant
gauge current to vanish at the horizon. Now, integrating the above equation on $y$ from $y_H$ to $y$, leads to
\begin{equation}
T_{\left(H\right)t}^{y}=\frac{1}{\sqrt{-g}}\left[b_{H}+\frac{e^{2}}{4\pi}
\left(A_{t}^{2}\left(y\right)+A_{t}^{2}\left(
y_{H}\right)\right)-\frac{e^{2}}{2\pi}A_{t}\left(y_{H}\right)
A_{t}\left(y\right)+N_{t}^{y}\left(y\right)-N_{t}^{y}\left(
y_{H}\right)\right]~,\label{eq30}
\end{equation}
where $b_{H}$ is an integration constant. On the other hand, $T_{t}^{y}\left(y\right)$ and $\nabla_{\mu}T_{t}^{\mu}$ can be written as
\begin{equation}
T_{t}^{y}\left(y\right)=T_{\left( o\right)t}^{y}\left(y\right)\Theta
\left(y-y_{H}-\epsilon \right)+T_{\left( H\right)t}^{y}\left(y\right)H\left(y\right)~,
\label{eq31}
\end{equation}
\begin{eqnarray}
\nabla_{\mu}T_{t}^{\mu}
&=&
\frac{1}{\sqrt{-g}}\left[-\frac{e^{2}}{2\pi}A_{t}\left(
y_{H}\right)\partial_{y}A_{t}\left(y\right)+\left(\sqrt{-g}
\left(T_{\left( o\right)t}^{y}\left(y\right)-T_{\left( H\right)t}^{y}\left(y\right)\right)+\frac{
e^{2}}{4\pi}A_{t}^{2}\left(y\right)+N_{t}^{y}\left(y\right)
\right)\delta\left(y-y_{H}-\epsilon\right)\right]\\
&+&\frac{1}{\sqrt{-g}}\partial_{y}\left(\left[\frac{e^{2}}{4\pi}A_{t}^{2}\left(
y\right)+N_{t}^{y}\left(y\right)\right]H\left(y\right)\right)~,
\label{eq32}
\end{eqnarray}
where we have used Eq.~(\ref{c0}). The first term in the above equation is a classical effect of the Lorentz force. The term in the total derivative is canceled
by quantum effects of classically irrelevant ingoing modes.
Therefore, the vanishing of the Ward identity under
diffeomorphism transformation implies that
\begin{equation}
T_{\left(o\right)t}^{y}-T_{\left(H\right)t}^{y}+\frac{1}{\sqrt{-g}}
\left(\frac{e^{2}}{4\pi}A_{t}^{2}\left(y\right)+N_{t}^{y}\left(
y\right)\right)=0~.\label{eq33}
\end{equation}
Then, by replacing Eq.~(\ref{eq28}) and Eq.~(\ref{eq30}) in the above equation, yields
\begin{equation}
a_{o}=b_{H}+\frac{e^{2}}{4\pi}A_{t}^{2}\left(y_{H}\right)
-N_{t}^{y}\left(y_{H}\right)~,\label{eq34}
\end{equation}
where $b_{H}$ is an integration constant, that has been chosen equal to zero in order for the covariant
energy-momentum tensor to vanish at the horizon. Therefore, the total flux of the energy-momentum tensor is given by
\begin{eqnarray}\label{energymomentumtensorflux}
\nonumber a_{o}&=&\frac{e^{2}}{4\pi}A_{t}^{2}\left(y_{H}\right)
-N_{t}^{y}\left(y_{H}\right)~,\\
&=&\frac{e^{2}}{2\pi y_{H}}Q^{2}q^{2}+\frac{1}{192\pi}f^{\prime}\left(
y_{H}\right)h^{\prime}\left(y_{H}\right)~.
\end{eqnarray}
Those results mean the flux required to cancel the gravitational anomaly at the horizon has a form equivalent to blackbody radiation, which we will show in the next section. Thus, the thermal flux required by black hole thermodynamics is able of canceling the anomaly.  

\section{Summary}

We studied the Hawking radiation from five dimensional dilatonic black hole, following the covariant anomalous point of view for gauge and gravitational anomalies proposed in Ref.~\cite{Gangopadhyay:2008zw}. The importance that had acquired the method of anomalies is that it let us to compute the Hawking temperature of the black holes. In the original version of the anomaly method, Ref.~\cite{Robinson:2005pd}, the authors considered scalar field, by simplicity. Near the horizon the theory can be described in two dimensions and it is possible to match the flux obtained via anomalies with the flux obtained from a beam of massless blackbody radiation moving in the positive r direction. Here, we considered a massive scalar field non minimally coupled to the curvature of this black hole background and it was shown that near the horizon, the physics can be described in two dimensions. We explicitly showed that the non-minimal coupling does not affect the dimensional reduction procedure, such as in Ref.~\cite{Papantonopoulos:2008wp}, since the part of the action which describes the non-minimal coupling vanishes in the mentioned region. Also, we considered Dirac fields in our description. In this case, we showed that dimensional reduction procedure can be developed. However, in order to match the fluxes, we consider the Hawking fluxes that can be obtained from integrating the Planck distribution for a general charged rotating black hole, for fermions.  The Planck distribution for blackbody radiation at the Hawking temperature  $T_H$, is given by
\begin{equation}
N_{e,m}(\omega)=\frac{1}{e^{\beta(\omega-e\Phi-m\Omega_H)}+1}.
\end{equation}
Then, the fluxes for the current and energy-momentum tensor can be computed by
\begin{equation}\label{eqlast}
F_Q=e\int_0^\infty\,\frac{d\omega}{2\pi}\left(N_{e,m}(\omega)-N_{-e,-m}(\omega)\right),
\end{equation}
\begin{equation}\label{eqlast}
F_M=\int_0^\infty\,\frac{d\omega}{2\pi}\,\omega\,\left(N_{e,m}(\omega)+N_{-e,-m}(\omega)\right),
\end{equation}
where we have included the contribution from the antiparticles. From the evaluation of Eq.~(\ref{eqlast}) for one charged and
static geometric we obtained $F_M=\frac{1}{4\pi}(e\Phi)^2+\frac{\pi}{12\beta^2}$, that exactly
match with Eq.~(\ref{energymomentumtensorflux}). Therefore, the Hawking temperature can be determined as
\begin{equation}
T=\frac{k}{2\pi}=\frac{1}{4\pi}\sqrt{f'\left(y_{H}\right)h'\left(y_{H}\right)}=\frac{1}{2\pi}Q\frac{\sqrt{m^{2}-q^{2}}}{m+\sqrt{m^{2}-q^{2}}}~,
\end{equation}
which coincides with the well-known result of the Hawking temperature for this geometry \cite{Nappi:1992as}. 

On the other hand, in order to avoid a general covariance and gauge invariance breakdown at the quantum level, the total flux in each outgoing partial wave of a quantum field must be equal to that of a (1+1) dimensional blackbody at the Hawking temperature with the appropriate chemical potential, while the total flux observed at infinity is that of a d-dimensional greybody at the Hawking temperature.


\section*{Acknowledgments}

We thank Joel Saavedra for his enlightening comments.

\appendix

\end{document}